# The Amenity Space and The Evolution of Neighborhoods

César A. Hidalgo[1,2] and Elisa E. Castañer[1],

[1] Macro Connections, The MIT Media Lab, Massachusetts Institute of Technology

[2] to whom correspondence should be addressed: hidalgo@mit.edu

**Abstract:**

Neighborhoods populated by amenities—such as restaurants, cafes, and libraries—are considered to be a key property of desirable cities. Yet, despite the global enthusiasm for amenity-rich neighborhoods, little is known about the empirical laws governing the colocation of amenities at the neighborhood scale. Here, we contribute to our understanding of the naturally occurring neighborhood-scale agglomerations of amenities observed in cities by using a dataset summarizing the precise location of millions of amenities. We use this dataset to build the network of co-location of amenities, or Amenity Space, by first introducing a clustering algorithm to identify neighborhoods, and then using the identified neighborhoods to map the probability that two amenities will be co-located in one of them. Finally, we use the Amenity Space to build a recommender system that identifies the amenities that are missing in a neighborhood given its current pattern of specialization. This opens the door for the construction of amenity recommendation algorithms that can be used to evaluate neighborhoods and inform their improvement and development.

## Introduction

How do businesses choose where to locate? For decades, scholars have been studying where businesses locate and why businesses agglomerate. But while the theoretical literature explaining the location and agglomeration of businesses is long and vast[1-10], the empirical literature documenting the location of business, especially at the intra-city scale, is much shorter and more recent.

The theoretical efforts explaining agglomerations in cities go back to Alfred Marshall's industrial districts[1], and to the mathematical models advanced by Johann Von Thünen[2], Harold Hotelling[3], Walter Christaller, and August Lösch[4-7]. In Von Thünen's model, differences in land use are explained as consequences of a location's distance from the market, which is a center of agglomeration[2]. In Hotelling's model, businesses agglomerate to maximize their catchment area—that is, to be the closest business to the largest number of potential customers[3]. In Christaller and Lösch's central place theory, a hierarchy of central places emerges when goods differ on how far a person would be willing to travel to purchase them[4-7].

In recent decades, however, these seminal models were expanded to endow them with economic micro foundations[8-10] and to include new mechanisms that could help explain agglomerations that were left out of the seminal models. Among these additional mechanisms we have demand externalities[11-12], which predict agglomerations when planned purchases trigger unplanned purchases; search costs, which predict agglomerations when customers like to compare prices[12]; and transportation effects, which predict agglomerations because transportation technologies reduce the cost of carrying goods and create an incentive to bundle purchasing trips[12].

The richness of this theoretical literature, however, has not been matched by equivalent empirical work, especially at the neighborhood scale. The relative scarcity of empirical work stems in part from the lack of data on the precise location of businesses and amenities, and in part because working at the neighborhood scale requires an implementable definition of what a neighborhood is. Not surprisingly, these limitations have pushed empirical work on agglomerations to focus on coarser scales, such as countries, cities, and regions, where data is readily available and the spatial units of analysis are exogenously defined[13-16].

Yet, scholars have still made important methodological progress at these coarser scales, by advancing network techniques to map products that are co-exported[13-14], or industries that hire similar workers[15-16]. The advantage of these network techniques is that they help preserve the identity of the elements involved in a dataset, and as such, are useful to study the effect of product and industry relatedness on the process of industrial diversification.

An example of this work in the context of economic development is the network connecting products that are likely to be co-exported, or product space[13-14], which has been used to anticipate the evolution of a country's export structure and the constraints that different productive structures impose on the ability of countries to generate income. Similarly, at the regional scale, scholars have used networks of similarity between industries and occupations to study the importance of skill relatedness in the evolution of regional economies[15-16].

Here we extend the use of these network methods to neighborhood scale agglomerations by using data on the precise location of more than 1.2 million amenities and by introducing a spatial clustering method that we use to identify neighborhoods. These data and methods allow us to solve the technical problems that have limited mapping the network of amenities that are likely to co-locate in the same neighborhood. We then validate the utility of this network of amenities, or Amenity Space, to build a recommender system that exploits a neighborhood's pattern of specialization to estimate the number of amenities of each type that should locate in it. This recommender system allows us to detect amenities that are potentially missing from a location, and also, represents an extension of the network methods used at the international and regional scale, to the neighborhood scale.

## Data

We use data from the Google Places API containing the latitude, longitude, and type of amenity (i.e. cafe, restaurant, library, etc.), for more than 1.26 million amenities across 47 US cities (see SM for details). Certainly the data from Google's Places API is not free of biases. The data on amenities registered in Google Places focuses on customer facing businesses and places of interests (from hair salons and bakeries to airports and cemeteries), and hence, fails to include information on other forms of economic activity, such as manufacturing. Also, the data might have coding issues, such as having a restaurant registered as a bar. Yet, despite these limitations, the Google Places API is accurate enough to be the backbone of the world's most popular mapping service (Google Maps) and is used daily by millions of individuals to find the location of businesses. This makes the Google Places API data an imperfect, yet attractive dataset to study the spatial organization of amenities at the intra-city scale.

Finally, we remind the reader that our results should be interpreted in the narrow context of the data from which these results were derived. This is data from an online mapping service and for U.S. cities only. The question of whether the results presented below can be generalized to other locations, and also, of whether these results hold for other datasets, is beyond the scope of this paper.

## Results

To identify what amenities co-locate in each neighborhood we begin by introducing a method to identify neighborhood scale agglomerations and the amenities that are present in each of them.

Our clustering procedure begins by calculating the *effective number of amenities* $A_i$ that are present in each location *i*. We define $A_i$ as the number of amenities that can be reached by walking from location *i*. Formally, the effective number of amenities in location *i* is a variant of what is known as an index of accessibility[17] $A_i$:

$$A_i = \sum_{j=1}^{N} e^{-\gamma d_{ij}}, \qquad (1)$$

where $d_{ij}$ is the distance between amenity *i* and amenity *j*, $\gamma$ is a decay parameter that discounts amenities based on their distance to location *i*, and *N* is the total number of amenities in city *c*. To interpret *A* it is useful to note that an amenity located where the measurement is taking place (i.e. with $d_{ii}$=0) contributes one to the effective number of amenities in that location, whereas an amenity at distance $d_{ij}$=1/γ—which would imply walking 1/γ kilometers from amenity *i* to *j*—will contribute only 1/*e* to that location's effective number of amenities ($A_i$). We find that our algorithm finds meaningful neighborhoods when we set γ=16, which implies that the contribution of an amenity to the effective number of amenities of a location roughly halves every 62.5 meters and becomes negligible at about 500 meters. This is consistent with research showing that the volume of pedestrian traffic becomes negligible after a ten-minute walk.[12,18]

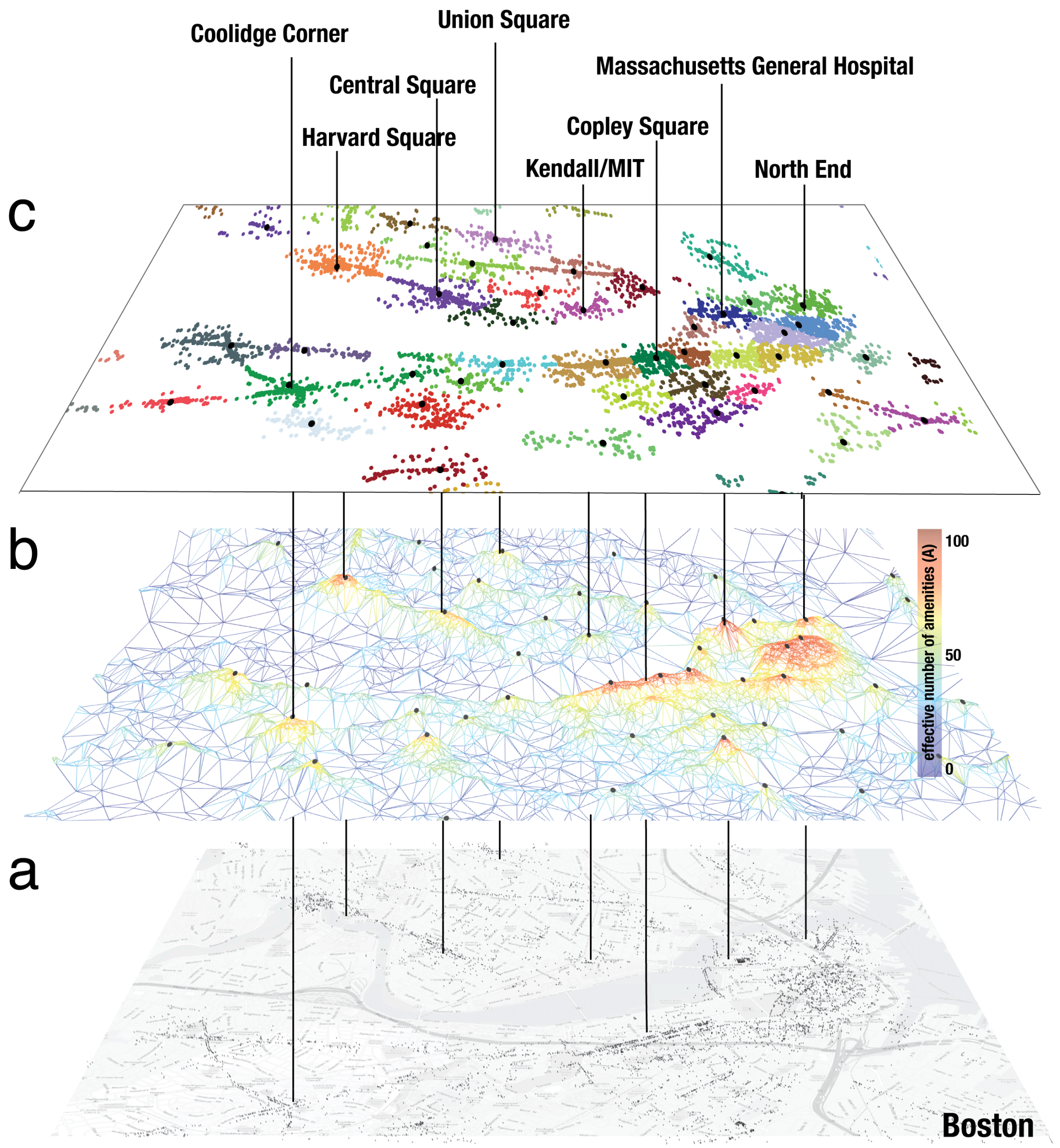

**Figure 1: Clustering algorithm. a** Map of Boston **b** The number of effective amenities ($A$) at each location where an amenity is present in Boston. Peaks represent locations with a high number of effective amenities and valleys represent locations with a low number of effective amenities. The black dots in the peak of the hills represent local maxima identified by our clustering algorithm, and are the center of neighborhoods. **C** Neighborhoods identified after the 90% of points with highest $A$ has been assigned to a location using our clustering algorithm. Neighborhoods are shown as sets of dots of the same color. Neighborhood centers are also marked by black dots.

We then use $A$ to identify the amenities belonging to each neighborhood using the following algorithm. First, we remove the 10% of amenities that have the lowest value of $A$, which represent amenities not located in an agglomeration. For the remaining 90% of amenities we identify local peaks on the landscape defined by $A$ (Fig 1b) by searching for locations that have an effective number of amenities that is larger than their $n$ nearest neighbors using the functional heuristic ($n_i = 3A_i + 50$). This heuristic helps avoid identifying multiple peaks in locations with a large concentration of amenities. Then, we assign amenities to each of the identified peaks using the following greedy algorithm: (i) we initialize each neighborhood by assigning to each peak all amenities that are in close proximity to it (less than 500 meters). Then, (ii) we calculate the distance between each amenity that has not been assigned to a neighborhood and all amenities that have been assigned to a neighborhood. Then, (iii) we assign to a neighborhood only the amenity that is closest to an amenity that has already been assigned to a neighborhood. Finally (iv), we recalculate the distance between assigned and unassigned amenities (repeat step (ii)) and assign one new amenity to a neighborhood by repeating step (iii). We continue until all amenities have been assigned to a neighborhood. An example of the neighborhoods found for the city of Boston is shown in Figure 1c (see SM for New York and San Francisco).

The neighborhoods identified using our algorithm (Fig 1c) correspond to well-known centers of urban activity. In the case of Boston these neighborhoods include Harvard Square and Central Square in Cambridge, and The North End and Coolidge Corner in Boston, among others.

We also note that the distribution of the effective number of amenities in a city is characterized by some universal properties. Figure 2a shows the distribution of the effective number of amenities ($A$) for every city in our dataset while Figure 2b shows the same distribution after normalizing the effective number of amenities in a city by that city's average effective number of amenities ($\langle A \rangle = \sum_i A_i / N$). For comparison, we show the same distributions but for an ensemble of cities where the location of each amenity has been randomized. These randomized cities are characterized by a narrow distribution, meaning that these random cities lack the high concentrations of amenities that signal the presence of neighborhood scale agglomerations in real cities. More importantly, figure 2b shows that once we normalize the effective number of amenities in a city by that city's average all cities follow the same lognormal distribution

$$P(\frac{A_i}{\langle A \rangle} = x) = \ln N\,(\mu, \sigma), \qquad (2)$$

with $\mu$ =-0.404 and $\sigma$=0.89. The existence of a universal distribution for the effective number of amenities across all cities in our sample means that all of these cities have an equal number of peaks and valleys of a given magnitude when the magnitude of these peaks and valleys is measured in units of that city's average.

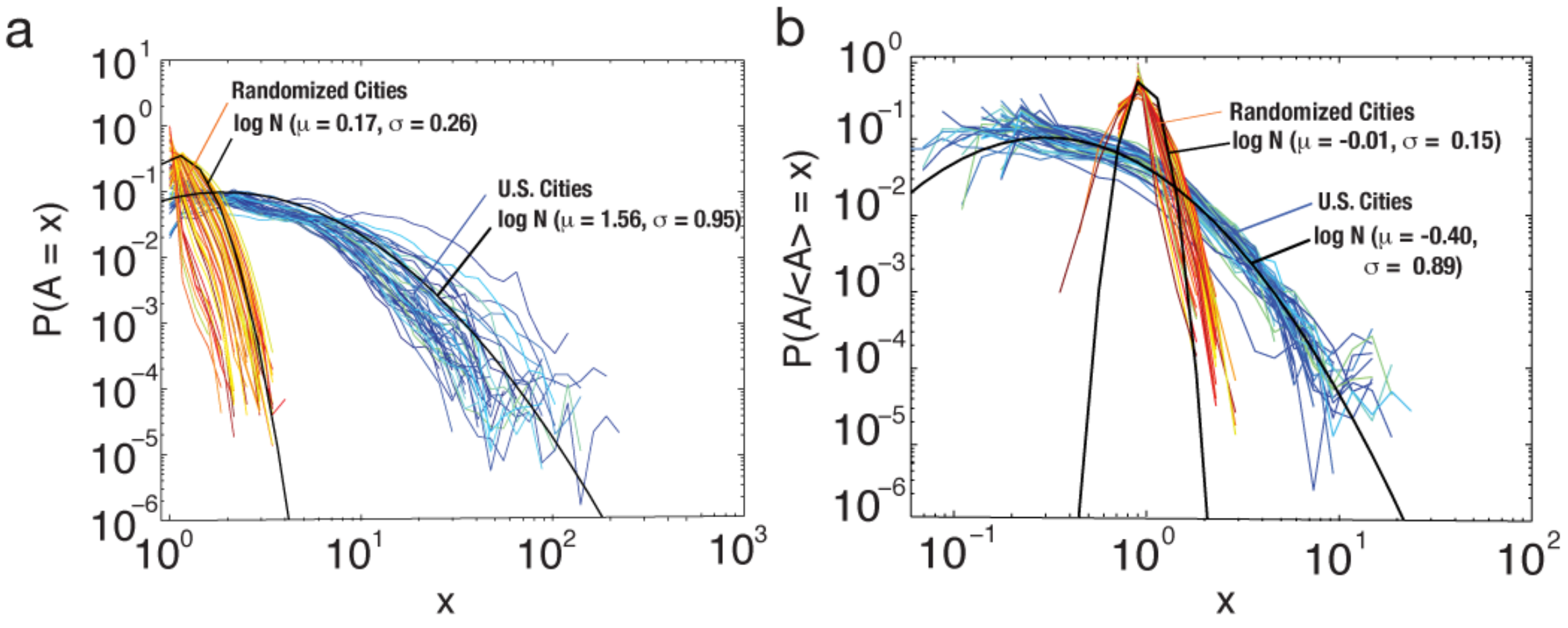

**Figure 2: City micro-agglomerations. a** The distribution of the effective number of amenities ($A$) in each US city. Blue lines show the distribution observed in our urban amenities data and orange lines show the distribution observed after randomizing the location of amenities for each city. **b** The distribution of the effective number of amenities ($A_i$) in each US city normalized by the average effective number of amenities in that city. Blue lines show the distribution observed in the cities data and orange lines show the distribution observed in the same cities but after randomizing the location of amenities

## The Amenity Space

After having identified neighborhoods for the 47 cities in our data we map the network connecting pairs of amenities that are likely to co-locate in the same neighborhood. We construct this network of amenities, or Amenity Space, by using spearman's rank correlation to identify pairs of amenities that are likely to be present in the same neighborhood. Figure 3a shows a visualization of this network containing the network's Maximum Spanning Tree[13] and the links that have a pairwise correlation equal or larger than 0.3 (see SM for the full correlations matrix). This subset of links provides a visualization that avoids visual clutter and that reveals what amenities tend to collocate with others. For example, car repair shops collocate with car dealers (Spearman's $\rho$=0.45), just like religious centers collocate with schools (Spearman's $\rho$=0.46). What is more important, however, is that this network tell us what combinations of amenities should predict the presence of others, providing a mean to recommend the amenities that should locate in the same neighborhood based on that neighborhood's current pattern of specialization. For instance, the network tells us that neighborhoods that specialize in beauty salons, accountants, and dentist, should also specialize in real state agents, but not in convenience stores or car rentals.

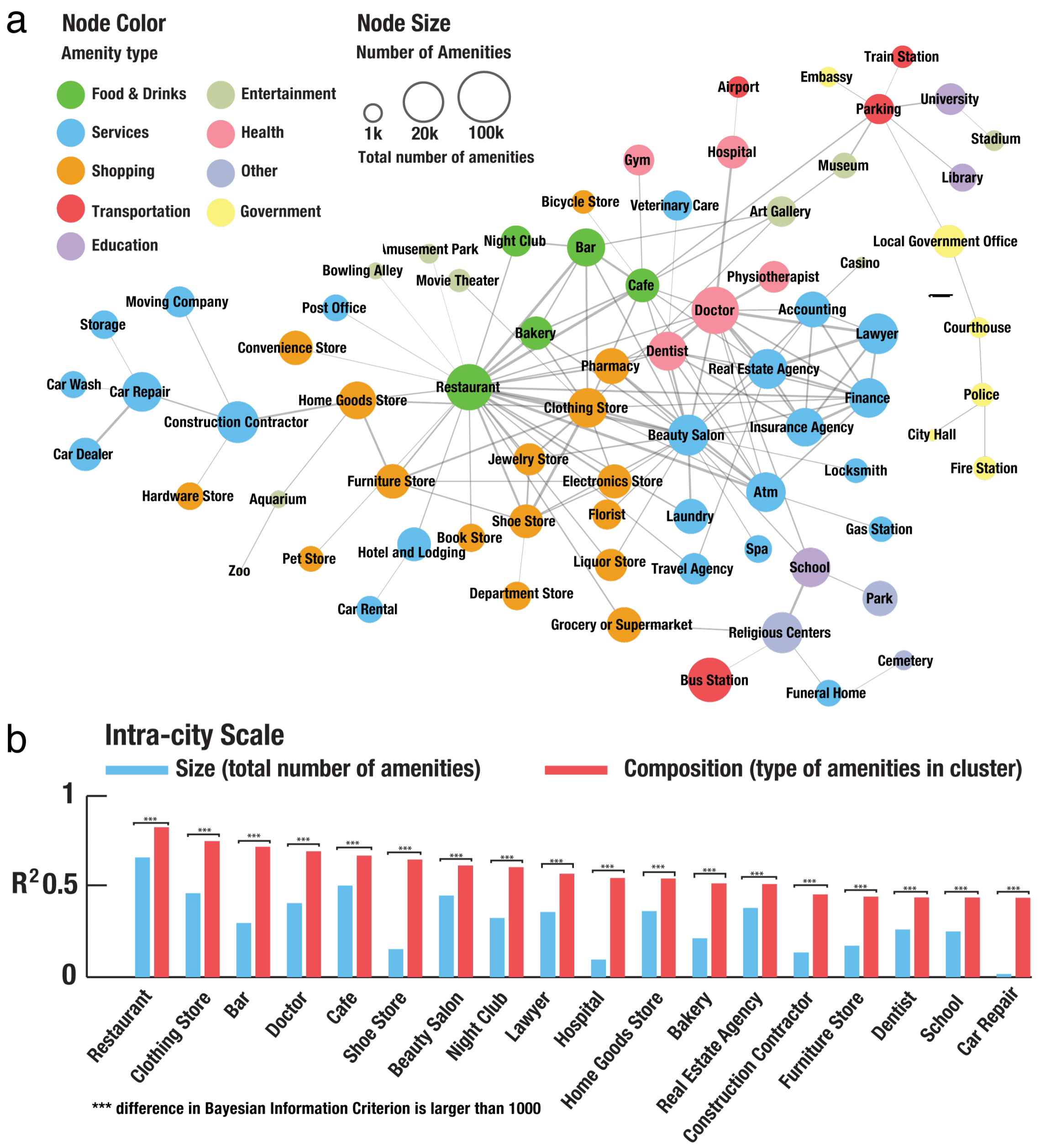

**Figure 3:** **a** Network of amenity co-locations. The nodes in the network represent different types of amenities and the edges connect amenities that are likely to collocate in the same neighborhood (see SM). The width of the edges connecting a pair of nodes is proportional to the spearman correlation obtained from the collocation of the two types of amenities across all neighborhoods. The size of a node is proportional to the number of times that an amenity is present in our data set. The color of each node represents the category that the amenity belongs to. **b** Comparison of the accuracy of two models used to predict the total number of amenities of each type on a neighborhood. The light-blue bars show the $R^2$ of a model predicting the number of amenities of each type in a neighborhood using only the total number of amenities in that neighborhood. The red bars show the $R^2$ of a model using information on the number of amenities of other types that are present in a neighborhood.

We then use the Amenity Space to build a parsimonious recommendation algorithm[19-20] for each type of amenity. We build this recommendation algorithm using multivariate regression and a forward selection algorithm that iteratively includes new types of amenities to the regression until the contribution of a new type of amenity is statistically insignificant (characterized by a *p*-value of more than 0.001 (see SM)). In addition, we control for over-fitting by using both Akaike's Information Criterion (AIC) and Bayes's Information Criterion (BIC). To provide a benchmark for the accuracy of the model we also predict the number of amenities of each type in a neighborhood using only the total number of amenities in that neighborhood (as a measure of the size of the economy of that neighborhood).

Figure 3b compares the $R^2$ of the models constructed using the patterns of specialization of neighborhoods with the models using only their size (the neighborhood's total number of amenities). In most cases (66/74=89%), the BIC test chooses the regression using the model based on the pattern of specialization over the regression using the neighborhood size (the exceptions are airports, aquariums, bus stations, car rentals, casinos, convenience stores, gas stations, and zoos), indicating that the model based on a neighborhood's pattern of specialization and the amenity space is better at predicting which amenity should locate in each neighborhood. Also, we note that the differences between the two models are not just statistically significant, but characterized by strong size effects. On average, for the 66 amenity types in which the amenity space model works better, the $R^2$ of the amenity space model is twice that of the model using size only ($R^2$=17% on average using size vs. $R^2$=35% on average using composition). This means that the increase in predictive power obtained by considering the types of amenities that locate in a neighborhood is not only statistically significant, but also substantial.

Finally, we illustrate our predictive algorithm by showing the recommendations it makes for specific amenities and neighborhoods for the city of Boston. Here, we recommend amenities by looking at the difference between the number of amenities observed in a neighborhood and those predicted by the model.

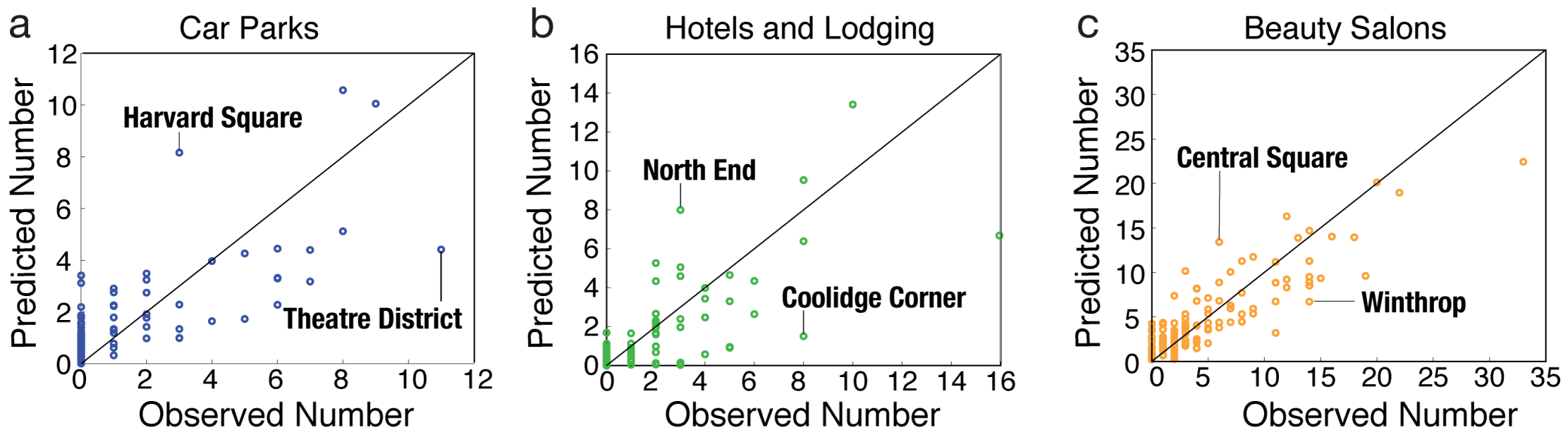

**Figure 5: Prediction of amenities in Boston's neighborhoods. a** Observed vs. predicted number of car parks, **b** hotels, and **c**, beauty salons for each neighborhood in Boston. Points above the lines represent neighborhoods where the predicted number of amenities is higher than the observed, suggesting instances of under supply. Points below the lines represent neighborhood where the predicted number of amenities is lower than the observed, suggesting instances of excess supply.

Figure 5 a-c compare the number of car parks, hotels, and beauty salons, observed and predicted, for each neighborhood in Boston. Points above the line, such as Harvard Square in the case of car parks (Figure 5a), the North End for hotels (Figure 5b), and Central Square for Beauty Salons (Figure 5c), indicate amenities that are under expressed in a location (given that location's current pattern of specialization). Points below the lines such as Boston's Theatre District in car parks, Coolidge Corner in hotels, and Winthrop in beauty salons, suggest instances of excess supply. Of course, the recommendations of the model should be taken with care. For instance, our model recommends more car parks in Harvard square, but of course, a decision to build new parking there should consider other aspects of Harvard Square that are not included in our model, such as the aesthetics of its architecture[21-22], or the externalities caused by cars. The lesson here is that the model successfully detects a known reality of Harvard square, which is that there is limited parking. Figure 5b shows another example in which our model detects a lack of hotels in the North End, a well-known tourist spot in Boston where only a handful of hotels are present. This could mean that there is a great potential for new hotels to locate in Boston's North End, but once again, that's a decision that would require additional considerations.

## Discussion

In this paper we introduced the Amenity Space, a network summarizing the patterns of co-locations characterizing neighborhood scale agglomerations, and used it to create a recommendation algorithm that we can use to evaluate the composition of amenities in a neighborhood.

Of course, our results and models are not free of biases and limitations. Beyond the data biases described above, our model is limited by its simplicity, which bounds the total amount of variance in the presence of amenities that we can explain. Our statistical model is based on linear regression techniques that could be potentially improved by using more complex functional forms, but also, by adding to them information that is not expressed in the presence of amenities, such as information on population density, the aesthetic appeal of a neighborhood's architecture[21-22], its foot traffic, and the daily and seasonal variations in traffic captured by mobile phone data[23]. Also, our model does not take into account zoning laws that can restrict the type of amenities that locate in each neighborhood.

Still, the results and methods presented here point to interesting new avenues of research. For example, time resolved data sources for both amenities and streetscapes could be used to explore the interaction between the dynamics of the amenities that locate in a neighborhood and the aesthetic of the buildings being constructed in it. Also, these results could help inform what types of business permits or incentives need to be given out to help balance a city's neighborhoods. On the computational side, the information uncovered here could also be used to create interactive online resources that would deliver the recommendations uncovered by our algorithm or similar algorithms. Together, our results, and the new avenue of research they open, should help stimulate the quantitative study of cities at the intra-city scale.

## Acknowledgements:

We acknowledge the support of The MIT Media Lab consortia and of Google's Living Labs award. We also acknowledge comments from Luis Bettencourt, Andres Sevtsuk, Scott Kominers, Mia Petkova, Jean-Francois Arvis, Luis Valenzuela, Matias Garretón, and Mark Hubberty.